\documentclass{elsart}

\usepackage{graphicx}
\usepackage{amssymb}

\begin{document}
\begin{frontmatter}

\title{On the potential of the imaging atmospheric Cherenkov technique
for study of the mass composition of primary cosmic radiation in
the energy region above 30 TeV}

\author[Heidelberg]{F.A. Aharonian},
\author[Barnaul]{V.V. Bugayov},
\author[Heidelberg]{J. Kettler},
\author[Barnaul]{A.V. Plyasheshnikov \corauthref{corr_author}}
\ead{plya@theory.dcn-asu.ru},
\author[Heidelberg]{H.J. V\"olk}

\corauth[corr_author]{Corresponding author.
Tel.: (+7)(3852) 36-70-75;
fax: (+7)(3852) 22-28-75}

\address[Barnaul]{Department of Physics, Altai State
University, Barnaul, Russia}
\address[Heidelberg]{Max-Planck-Institut f\"ur Kernphysik,
Heidelberg, Germany}
\begin{abstract}

The technique of imaging atmospheric Cherenkov telescopes (IACTs)
has proved to be an effective tool to register cosmic
$\gamma$-rays in the very high energy region. The high detection
rate of the IACT technique and its capability to reconstruct
accurately the air shower parameters make it attractive to use
this technique for the study of the mass composition of cosmic
rays. In this article we suggest a new approach to study the CR
mass composition  in the energy region from 30 TeV/nucleus up to
the "knee" region, i.e. up to a few PeV/nucleus, using an array of
imaging telescopes  of a special architecture. This array consists
of telescopes with a relatively small mirror size ($\sim
10$~m$^2$) separated from each other by large distances ($\sim
500$~m) and equipped by multichannel cameras with a modest pixel
size ($0.3-0.5^{\circ}$) and a  sufficiently large viewing angle
($6-7^{\circ}$).

Compared to traditional IACT systems (like HEGRA, HESS or VERITAS) the
IACT array considered in this study could provide a very large
detection area (several km$^2$ or more). At the same time, it allows an
accurate measurement of the energy of CR induced air showers (the
energy resolution ranges within 25-35\%) and an effective separation of
air showers created by different nuclei. Particularly, it is possible
to enrich air showers belonging to the nucleus group assigned for
selection up to $\sim 90\%$ purity at a detection efficiency of 15-20\%
of such showers.
\end{abstract}

\end{frontmatter}

\section{Introduction}

Last years the technique of imaging atmospheric Cherenkov
telescopes (IACTs) has proved to be an effective tool to register
cosmic $\gamma$-rays in the energy region above 100~GeV
\cite{Cawley,Hillas,Aharonian1}. In comparison to the satellite
experiments this technique provides as much as a few magnitudes
larger detection area. Compared to the particle detector arrays it
comprises a low energy threshold of detected  air showers. In
addition, IACT arrays  (like HEGRA \cite{HEGRA}, HESS \cite{HESS}
or VERITAS \cite{VERITAS}) provide the ability  of accurate
reconstruction of air shower parameters such as the arrival
direction, the core location and the primary energy {\it both} for
$\gamma$-ray and cosmic ray (CR) induced air showers. All this
makes it  attractive  to apply the IACT technique for the study of
the CR mass composition.

In \cite{Plyasheshnikov2} an approach was developed allowing to
use the IACTs for study of the chemical composition of CRs in the
primary energy region $\ge$1~TeV/nucleus. Later this approach
received an application in the analysis of the observational data
of the HEGRA collaboration (see Ref.~\cite{PhysRev}). In this
study we also discuss possibilities of application of the IACT
technique for an analysis of the CR mass composition. However, on
the contrary to \cite{Plyasheshnikov2}, we analyse here higher
primary energies, i.e. the energy interval between a few dozen
TeV/nucleus and the "knee" region where the energy spectrum
steepens from differential spectral index $\sim 2.7$ below $\sim
3$ PeV to $\sim 3.0$ above (see Fig.~1). This energy
region becomes important for ground based observations hence the
balloon and satellite born experiments run out of statistics here.
The region around the "knee" is of special interest, because in
all likelihood it contains contributions of different types of CR
sources. Presumably this could be the case when the contribution
to CR population made by the supernovae remnants of the Galaxy
changes for sources of other origin (see e.g.
Ref.~\cite{Shibata}). Different types of sources and a rather
smooth energy dependence in the "knee" region combined with the
varying properties of interstellar medium permits sufficient
freedom in selection of alternative acceleration models. In such a
circumstance it is important to develop new methods enabling an
accurate determination of CR chemical composition and capable of
reconstructing the structure of energy spectra in the "knee"
region for individual CR nucleus groups.

A new approach to the analysis of the CR mass composition
developed in this study is based on a special kind of IACT array
consisting of telescopes separated by distances considerably
larger than for traditional IACT systems and equipped  by
multichannel cameras with a wide field of view\footnote{An idea to
apply an IACT array with such an architecture for observations of
$\gamma$-ray sources in the multi-TeV energy region was first
suggested in \cite{Plyasheshnikov1}.}. We show that such an array
could provide a huge detection area (up to several km$^2$) and,
thus, could extend an interval of observed primary energies up to
the "knee" region. A detailed study is undertaken to estimate the
possibility of the array to measure the energy of individual air
showers and to separate showers induced by different primary
nuclei in the "event by event" mode, which is the natural way of
data analysis for stereoscopic IACT observations. Besides, such an
approach, being a realization of the so-called non parametric
analysis \cite{Chilingarian}, yields not only an estimate of the
primary energy and mass composition, but also allows to specify
the uncertainty of the results in a quantitative way.

\section{Simulations}
All results presented in this paper have been obtained by means of the
ALTAI simulation code \cite{Konopelko1} developed particularly for the
simulations of Cherenkov light emission produced by $\gamma$-ray and
proton/nuclei-induced air showers. Being essentially less time
consuming than the CORSIKA code \cite{CORSIKA}, ALTAI well reproduces
experimental data obtained by the HEGRA and Whipple collaborations (see
Ref. \cite{ALTAI_HEGRA} and \cite{ALTAI_WHIPPLE} respectively). Air
showers created by four different types of primaries (protons,
$\alpha$-particles, oxygen and iron nuclei) have been considered for
the primary energy region up to $10^3$~TeV/nucleus.

We carried out our simulations for a system of IACTs located at the
corners of a square considering different values of the square side
length $l$.  The optical axes of the IACTs were assumed to be parallel
to each other and vertically directed. Following \cite{Plyasheshnikov1}
we consider this system of IACTs as a cell  of an IACT array consisting
of a large number of cells ($K\gg 1$) and having a detection area
$S\simeq K\cdot l^2$. In this relation, we assume that the core of air
showers registered by the cell is uniformly distributed inside the cell
square\footnote{If an IACT array has a number of cells $K\gg 1$, this
assumption is valid for an overwhelming majority of registered air
showers.}.

The IACT considered in our analysis  has a modest mirror size (10
m$^2$) and an ordinary value ($\sim$0.1) of the Cherenkov
photon/photoelectron conversion  factor. A number of hexagonal
configurations of the multichannel camera has been  considered for this
telescope with the field of view (FoV) up to  $\sim 9^o$. Two values of
the camera pixel size ($PS=0.3^\circ$ and $0.5^\circ$) have been
studied.

We assume that the cell registers air showers triggering simultaneously
at least two of its telescopes and apply the following condition  to
simulate the hardware trigger of an individual  telescope: at least two
adjacent pixel magnitudes of the  central part of the camera (with $N$
pixels) should exceed a  given value $q_0$. The minimum value of $q_0$
is determined by the  night sky light contribution, the value of
parameter $N$ and the number ($L$) of telescopes triggered
simultaneously (see for details \cite{Plyasheshnikov1}). In the case of
$L\ge 2$ we  have $q_{0} \sim 10$ photoelectrons  for $PS=0.3^o$ and
$q_{0} \sim 20$~ph.e. for $PS=0.5^o$.

To estimate the IACT detection rate one needs an input about the CR
energy spectra. Strictly
speaking, these spectra differ from each other for different nucleus
groups (see e.g. Ref.~\cite{Wiebel}). We neglect for simplicity this
difference and assume that CR nucleus groups
give an energy-independent contribution to the all-particle CR flux
\begin{equation}
dF_{CR}/dE = 0.25\cdot E^{-2.7} (\mathrm{s \, sr \, m^2 \, TeV})^{-1}
\label{Spectrum}
\end{equation}
The data on this  contribution are presented in
Table~\ref{Assumed}. Each value given in the table is the
proportion of the respective nuclei group in the all-particle
flux.

\begin{table}
\caption{The assumed chemical composition of primary cosmic rays.}
\begin{tabular}{|l|c|c|c|c|} \hline
Nucleus group                          &    P&  $\alpha$&     LM&
HVH  \\ \hline Atomic number range                    &    1& 4&
5-20& 21-56  \\ \hline Proportion in the primary CR flux & 0.41&
0.27&   0.12& 0.20  \\ \hline
\end{tabular}
\label{Assumed}
\end{table}

The results presented here correspond to either a fixed value of
the energy of CR protons and nuclei or a power-law energy spectrum
of these particles calculated in accordance with
Eq.~(\ref{Spectrum}) and Table~\ref{Assumed}.

In our study we consider the IACT detection rates corresponding to
four
groups of CR primary particles -- protons, $\alpha$-particles,
'low + medium' (LM) and 'heavy + very heavy' (HVH) nuclei.
For LM and HVH groups we  assume for simplicity
that all air showers of these groups are  induced by nuclei with the
same value of the atomic number (oxygen and iron nuclei,
respectively).

To evaluate parameters of the Cherenkov light image we apply a
traditional second-moment approach (see e.g. Ref.~\cite{Hillas2}).
By rotation of the reference frame it is always possible to choose
a new coordinate system in which the matrix of second moments is
diagonalized. The diagonal elements equal to $Length^2$ and
$Width^2$ characterize the extent of the Cherenkov light
distribution along the major and minor axes of the image and are
used by us further in the paper. To ensure a high quality of
images and to reduce the influence of the night sky background a
two level cleaning procedure is applied. At the first step of this
procedure those pixels are included in the image which contain not
less than 5 ph.e. ($PS=0.3^{\circ}$) or 9 ph.e.
($PS=0.5^{\circ}$). At the second step those pixels are excluded
that neither contain 7 (12 for $PS=0.5^{\circ}$) or greater ph.e.
nor have in the neighbourhood pixels with a nonzero content.

\section{Layout of the cell and configuration of the camera}

In this section we discuss problems connected with an
optimal choice of the layout of the IACT cell (the sidelength $l$
of the square) and the  configuration of the
multichannel camera, i.e.
the field of view ($FoV$) and  the pixel size ($PS$).

In Fig.~2 we present the total number of photoelectrons
in the Cherenkov light image
(the $Size$ parameter according to notification
of Ref.~\cite{Hillas2}) as a function of the air
shower impact parameter $r$\footnote{We define the impact parameter
as the distance between the
telescope and the location of the air shower core at
the observation level}. Data of the figure
correspond to different values of the primary energy and different
types of primary particles.
Assuming a minimum acceptable value of $Size$ ($\sim$ 100~ph.e.)
to provide a sufficiently good
quality of the image, one can see that in the energy
region $\ge 10$~TeV an IACT with a modest
mirror size (10 m$^2$ in the case of Fig.~2) could
detect CR induced air showers with
very large values of the impact parameter (up to $r \sim 500$~m).
For energies higher than
$\sim$~100~TeV impact parameters of registered air showers are
extended up to $\simeq 10^3$~m.
Moreover, in the energy region $\ge 100$~TeV even an IACT with
mirror size $\simeq 1
\mathrm{m}^2$ could provide an effective registration of air showers
from distances up to $\sim 500$~m.

Keeping in mind the conclusion made in the previous paragraph, the fact
that a multitelescope IACT array consisting of a big number ($K$) of
cells has the detection area $S\simeq K\cdot l^2$ and the fact that the
integral CR flux $F(\ge E)$ decreases rapidly with $E$ one can reach
the following conclusion: to study the CR properties by the IACT
technique in the energy region considered in this study it is
reasonable to choose the distance between telescopes to be considerably
larger than for the traditional IACT arrays (e.g. $\sim 85$~m for the
HEGRA array).

In Fig.~3 we present the energy dependence of the
differential detection rate of a cell for two sufficiently large values
of sidelength ($l=$333~m and $l=$500~m) and four different groups of
primary nuclei. The data in the figure correspond to the simultaneous
triggering of two or more IACTs of the cell. The minimum available
value of the hardware trigger parameter $q_0=10$ is used. It is seen
from Fig.~3 that the energy dependence of the differential
detection rate has a local maximum. The position of this maximum  can
be considered as an effective energy threshold $E_{\mathrm{th}}$ of the
IACT cell. The values of $E_{\mathrm{th}}$ are presented in
Table~\ref{Thresholds}.

\begin{table}
\caption{The effective energy threshold (TeV) of the IACT cell for
different layouts and different primary particles. $PS=0.3^o$,
$FoV=6.3^o$, $q_0=$10~ph.e.}
\begin{tabular}{|l|c|c|c|c|} \hline
Primary nucleus                  &   P&  He&     O&      Fe  \\
\hline $l=$333~m         &   7&   8&    12&      20   \\ \hline
$l=$500~m         &  16&  22&    31&      38   \\ \hline
\end{tabular}
\label{Thresholds}
\end{table}

The  following conclusions can be obtained from Fig.~3
and Table~\ref{Thresholds}.

\begin{itemize}
\item
Both values of $l$ provide an   effective energy threshold
belonging
to the multi-TeV region. The value of this threshold increases
with the
atomic number  ($A$) of the primary nucleus and ranges within
$7-20$~TeV
for $l=$333~m and $16-38$~TeV for $l=500$~m.
\item
Dependence $E_{\mathrm{th}}(A)$ is  considerably weaker than for
the case of traditional IACT systems. In our case difference
between $E_{\mathrm{th}}$ corresponding to proton and the iron
nucleus is $\sim 2\div 3$, whereas for traditional IACT system
this difference can reach a factor of 8
\cite{Plyasheshnikov2}\footnote{Registration of air showers with
small values of the impact parameter ($r\le 100$~m) predominates
in the case of traditional IACT arrays. In this region of $r$ the
$Size$ parameter (see Fig.~2) reduces more rapidly with
$A$ than at large $r$.  As a result  a more considerable growth of
$E_{\mathrm{th}}(A)$ is observed.}.
\end{itemize}

For different telescopes of an IACT array the location of images of
detected air showers in the focal plane of the telescope mirror is
different, because this location correlates considerably with the value
of the shower impact parameter\footnote{Variation of the impact
parameter as much as 100~m results, in average (see e.g.
Ref.~\cite{Plyasheshnikov1}), in $\simeq 1^o$  displacement of the
image along the focal plane.}. In this relation, an effective
registration of air  showers  by a cell with large value of $l$ (when
two or more telescopes are triggered  simultaneously) can be reached
only with wide angle multichannel cameras. One can check this statement
inspecting Fig.~4 where we present the ratio of the cell
detection rate ($R$)  to the field of view ($\Delta\Omega$) of the
camera as a function of the viewing angle $\theta$. It is seen that to
keep the ratio $\Delta=R/\Delta\Omega$ near its asymptotic value
$\Delta_{\infty}=\Delta(\theta\to \infty)$ one should use a camera with
the viewing angle larger than a definite value $\theta_0$. Assuming,
for example, $\Delta\simeq 0.5\cdot \Delta_{\infty}$ we have
$\theta_0\sim 5-6^o$ for $l=$500~m and $\theta_0\sim 2-3^o$ for
$l=333$.

Noted above values of $\theta_0$ correspond  to the hardware
trigger condition  $2/N_0\ge q_o$, where $N_0$ is the total number
of camera pixels. This condition pays no attention to an accurate
determination  of parameters of those images that lie near the
edge of the camera.  To avoid distortion of such images one should
apply, instead of $2/N_0\ge q_0$, a trigger condition involving
only the central part of the  camera with a $FoV$ less than the
total $FoV$ by as much as the angular  size of the image. Assuming
for the image size the value $\sim 1^o$ (see section~5), we have
for the full viewing angle of the camera the following
restrictions: $\theta\ge 3-4^o$ for $l=333$~m  and $\theta\ge
6-7^o$ for $l=500$m.

As it is shown in \cite{Plyasheshnikov1} for the cell with $l=500$~m
the optimum value of the pixel size ($PS$) is about $0.3^o$, if one
detects $\gamma$-ray induced air showers. The angular size of CR
induced images is larger than that for $\gamma$-rays. Therefore, for
our purpose there are no reasons to use $PS\le 0.3^o$. For the case of
$PS=0.3^o$ the total number of channels of camera with the optimum
field of view is  not very large. For example, $N_0=469$ for
$FoV=7.5^o$ (see Table~\ref{NPix}).

\begin{table}
\caption{Relation between the FoV of a hexagonal camera and the
total number of its pixels.}
\begin{tabular}{|l|c|c|c|c|c|} \hline
$PS=0.3^{\circ}$& Pixel  number&    271&  331&    397& 469  \\
\hline $PS=0.3^{\circ}$& $FoV$, degree&    5.7&  6.3&    6.9& 7.5
\\ \hline $PS=0.5^{\circ}$& Pixel  number&    91&  127&    169&
217  \\ \hline $PS=0.5^{\circ}$& $FoV$, degree&    5.5&  6.5& 7.5&
8.5\\ \hline
\end{tabular}
\label{NPix}
\end{table}

In addition to $PS=0.3^o$  a larger value of pixel size ($0.5^o$) has
been analyzed by us.  This analysis has shown that there exists no
dramatic difference between the basic properties of the cell  (the
energy  resolution, the error of determination  of the shower core
location, the ability to separate air  showers  initiated  by different
primary nuclei) corresponding to such values of $PS$. At the same time,
for $PS=0.5^o$ the total number of camera pixels (see Table~\ref{NPix})
is smaller by factor of $\sim 3$ compared to $PS=0.3^o$.

Let us now formulate conclusions showing that the cell layout with
$l=500$~m looks to be close to an optimum one.

\begin{itemize}
\item
This layout provides an effective energy threshold
close to the lower boundary of the energy region
considered here.
\item
A multitelescope  array with this value of $l$ provides a sufficiently
large value of detection area ($\sim$ 0.25~km$^2$ per one cell).
\item
For this cell one has to use a multichannel camera with a rather
wide viewing angle ($\theta\sim 6-7^o$). However, due to a
sufficiently large value of the optimum pixel size
($0.3-0.5^{\circ}$), the total channel number for this camera is
not very large.
\end{itemize}

Let us now list the basic parameters of the IACT cell considered by us
everywhere in further analysis.  Such a cell has the sidelength
$l=$500~m. The cameras of the cell telescopes have the viewing angle
$7.5^{\circ}$ and the pixel size $0.3^{\circ}$ (or $0.5^o$). The total
number of the camera channels is 469 (169 for $PS=0.5^o$). The air
showers detected by the cell trigger simultaneously at least two of its
telescopes. The following condition is used for the hardware triggering
of individual telescopes: $2/N\ge q_0$ where $N=331$ (127 for $0.5^o$)
is the pixel number of the central part of the camera and
$q_0=10$~ph.e. (20 ph.e. for $PS=0.5^o$).

\section{Statistics of detected events}
In Table~\ref{Props} we present the detection rates of the IACT cell
with a layout and camera configuration close to the optimum ones for
different groups of nuclei. Besides, we show in this table the
proportions of these nucleus groups in the total detection rate of the
cell. It is seen that the proportion of heavy nuclei in the detection
rate is somewhat lower than this proportion in the primary CR flux
(compare Table~\ref{Assumed} and Table~\ref{Props}). This peculiarity
can be explained by the following way. A growth of the effective energy
threshold $E_{\mathrm{th}}$ takes place with an increase of the atomic
number of the primary nucleus (see Table~\ref{Thresholds}); as a result
heavy nuclei are registered less effectively compared with light ones.
At the same time, the proportion of HVH nuclei in the cell detection
rate ($\sim$ 15\%) is considerably larger than this quantity for
traditional IACT systems (less than 5\% according to
Ref.~\cite{Plyasheshnikov2}). This circumstance is connected with the
fact that in the latter case an substantially stronger dependence of
$E_{\mathrm{th}}$ on the atomic number is observed. This effect has
already been discussed in section~3.

\begin{table}[ht]
\caption {The detection rates ($R_{\beta}$) for different nucleus
groups ($\beta$) and their proportion ($\delta_{\beta}$) in the
total rate  ($R=\sum R_{\beta}$) of the cell.}
\begin{tabular}{|l|c|c|c|c|}\hline
Nucleus group, $\beta$     &P    &He   &LM   &HVH  \\ \hline
$\delta_{\beta}$          &0.49 &0.26  &0.10  &0.15  \\ \hline
$R_{\beta}$, event/hour   &1000 &543   &199   &311    \\ \hline
\end{tabular}
\label{Props}
\end{table}

The proportions $\delta_{\beta}(\ge E)$ of different nucleus
groups $\beta$ as functions of the lower energy bound have also
been considered in this study. Analysing these quantities we come
to a conclusion that the energy dependence of quantities
$\delta_{\beta}(\ge E)$ is quite weak. For the whole energy
interval 30-1000~TeV considered here CR protons are the most
numerous in the detection rates, whereas  the LM nuclei are the
least numerous.

Selection criteria aimed at the separation of a fixed nucleus
group reduce the detection rate of this group\footnote{Typically
by a factor of $5-10$ according to section~7.}. Nevertheless, even
after the separation of nuclei the technique considered here could
provide a sufficiently good statistics of observations.
One can check this statement inspecting Table~\ref{Statistics}.
In this table we present the calculational results on the number
of protons and HVH nuclei detected by an IACT array with a modest
number of cells (16) and passed through a software  separation
procedure providing for selected nuclei a  typical value of the
acceptance probability $\kappa=0.15$. A set of values of the
triggering parameter $q_0$ is used corresponding to different
values of the effective energy threshold of the array. It is seen,
for example, that  about 2000 HVH nuclei could be detected in the
energy region above $\sim 200$~TeV for a modest observation time
(100~h). This number is  essentially greater than the total number
of such nuclei have been accumulated until now in the balloon and
satellite born experiments \cite{Shibata}.

\begin{table}
\caption{The mean numbers $\bar{N}$ of CR protons and HVH nuclei
which could be detected by an IACT array consisting of 25
telescopes during 100~h of observations after a separation
procedure saving $\sim$15\% of nuclei assigned for selection.
$E_{\mathrm{th}}$ is   the effective energy  threshold (TeV) of
the cell.}
\begin{tabular}{|l|c|c|c|c|c|c|c|c|c|}  \hline
  &\multicolumn{2}{c|}{P}& \multicolumn{2 }{c|}{HVH}&
  &\multicolumn{2}{c|}{P}& \multicolumn{2 }{c|}{HVH}
  \\ \hline
 $q_0$& $E_{\mathrm{th}}$& $\bar{N}$& $E_{\mathrm{th}}$& $\bar{N}$&
 $q_0$& $E_{\mathrm{th}}$& $\bar{N}$& $E_{\mathrm{th}}$& $\bar{N}$
\\ \hline
    10&       16& $9.3\cdot 10^4$&   38& $5.3\cdot 10^4$&
    30&       61& $1.9\cdot 10^4$&   141& $8.0\cdot 10^3$ \\  \hline
    20&       43& $2.9\cdot 10^4$&   101& $1.3\cdot 10^4$&
    60&      149& $4.3\cdot 10^3$&   203& $1.9\cdot 10^3$ \\  \hline
\end{tabular}
\label{Statistics}
\end{table}

\section{Image parameters}
In Fig.~5  we present the probability distributions of the
impact parameter ($r$) of air showers with respect to each triggered
telescope of the cell with different values of the sidelength $l$. It
is seen that the shape of the probability distribution of $r$ depends
rather weakly on the type of the primary nucleus. At the same time,
there exists a considerable dependence of this distribution on the
$l$-parameter. For example, for $l=$333~m the most probable value of
$r$ $\simeq$200~m, whereas for $l=500$~m it is $\simeq 300$~m. For both
values of $l$ the most probable value of $r$ essentially exceeds this
quantity for traditional IACT arrays ($\le 100$~m).

In Fig.~6 the mean values of the basic shape parameters
($Width$ and $Length$) for primary proton are presented as
functions of $r$ for different intervals of the primary energy. It
is seen that the mean value of $Width$ reduces with $r$, whereas
for the $Length$ a growth of the mean value takes place. As a
result CR-induced images detected by the cell (in comparison to
traditional IACT systems) have more elongated shape with a well
established orientation\footnote{For $\gamma$-ray induced air
showers this effect was first noted in
Ref.~\cite{Plyasheshnikov1}.}. An increase of the primary energy
results in a growth of both shape parameters (by a factor of
1.5-2.0 in the energy region 30-1000~TeV).

An effective separation  of air showers created by different
primary nuclei by the IACT technique is possible only, if there
exists a considerable sensitivity of image parameters to the
variation of the atomic number $A$. The $Width$ parameter exhibits
a strong correlation with $A$ (see Fig.~7).
In particular, the most probable value of this parameter increases
from $0.18^o$ for protons to $0.32^o$ for iron nuclei. Dependence
on $A$ of the second shape parameter ($Length$) is quite weak; its
most probable value ranges within $0.6-0.7^o$.

In Fig.~8 we present the probability distributions of one
more parameter
relatively sensitive to $A$. This is the altitude $H_{\mathrm{max}}$
at which the maximum amount of the Cherenkov light is
emitted\footnote{The value of $H_{\mathrm{max}}$ can be
reconstructed on the basis of the location
of the shower core at the observation level and location
of the "center of gravity" of the image in the focal plane of the
telescope mirror (see e.g. Ref.~\cite{Hmax}).}. Besides,
our analysis has shown that the $Conc$ parameter defined as
the ratio of two largest pixel magnitudes to the total number of image
photoelectrons is also sensitive to the variation of
$A$\footnote{However this parameter correlates considerably
with $Width$.}.

\section{Air shower parameters}
One important feature  of the IACT technique is the possibility to
determine for individual air showers their basic parameters such
as the core location, arrival direction and primary energy. In
this section we discuss the accuracy of the determination of these
parameters by the cell with optimum layout and configuration of
the multichannel camera. Reconstructing these parameters we follow
an approach described in \cite{Aharonian3}.

We define the core location error for an individual air  shower as
$\delta r=((x-x_0)^2+(y-y_0)^2)^{1/2}$ where $x,x_0,y,y_0$ are the
Cartesian  coordinates (at the observation  level) of the real
($x,y$) and measured ($x_0,y_0$) core location.  The curves in
Fig.~9 present the order ($\delta r_0$) of the core
location error defined by the following way: an inequality $\delta
r \le \delta r_0$ should be satisfied for 50\% of detected air
showers. One can see from Fig.~9 that at high energies
$\delta r_0$ depends rather weakly on both primary energy and
atomic number $A$. It is approximately equal to 30~m in this
energy region. Reduction of the primary energy leads to a growth
of the error and strengthens its dependence on $A$. For example,
at $E=100$~TeV the error $\delta r_0$ ranges between $\simeq 30$~m
for the proton and $\simeq$ 60~m for the iron nucleus.

We define the energy resolution $\delta E_0$ of the cell by the
following way: for 50\% of detected air showers an inequality $\delta E
\le \delta E_0$ should be satisfied, where $\delta E$ is the relative
error of energy determination for an individual air shower. The
calculations show that this resolution ranges within $25-35$\% for the
entire considered energy region. It becomes worse with a growth of the
primary energy and depends slightly on the atomic number of the primary
nucleus (see Fig.~10). In Fig.~11 we compare results
on $\delta E_0$ for two different values of the camera pixel size
($0.3^o$ and $0.5^o$) with regarding (or disabling) the optical
aberrations of the IACT mirror\footnote{We use an approach developed in
Ref.~\cite{Ashot} to describe optical aberrations.}.
It is seen that both these effects can not seriously influence the
quantity $\delta E_0$.

Thus, the approach considered here provides
a sufficiently good energy resolution allowing a precise
determination of the CR energy spectra for different
nucleus groups. In particular,  the value of $\delta E_0$ is much
smaller compared  to the scale of the CR ``knee'' region.

\section{Separation of air showers}
Selection of a given nucleus group is essentially a classification
problem. For solution of this problem one needs to construct a
rule which connects given values of image parameters with the type
of primary particle. This rule allows to decide whether a given
set of parameters is peculiar for the nucleus group assigned for
selection. The rule can be expressed in terms of a function of $N$
arguments, each of them corresponds to a specific image parameter.
If, for example, this function is smaller than a definite value,
then it is most likely that this image is induced by the nucleus
group assigned for selection. Otherwise, the event should be
rejected. In this study we adopt the function which is similar to
distance in multi-dimensional parameter space. A criterion based
on such a function was applied, in particular, for rejection of
the cosmic ray background in Ref.~\cite{ESpec}. It is called there
"ellipsoidal-window gamma-ray image selection criterion". The
distance is defined by the following formula:
\begin{equation}
d^2=\sum_{i=1}^{N}{\sum_{j=1}^{N}{\mathbf{M}_{ij}
(p_i-\bar{p_i})(p_j-\bar{p_j})}} \label{d2M}
\end{equation}
where $p_i (i=1,...,N$) are the image parameters, $\mathbf{M}_{ij}$
is the matrix which define the metrics of parameter space in this
specific approach and calculated as the inverse of the second
moment matrix $\mathbf{M}_{ij}=\langle
(p_i-\bar{p_i})(p_j-\bar{p_j})\rangle^{-1}$. Using an appropriate
value $\tilde d^2$ of this parameter one can obtain a decisive
rule in the form $d^2<\tilde d^2$.

In our approach we apply the function which is slightly different
from Eq.~(\ref{d2M}):
\begin{equation}
d'^2=\sum_{i=1}^{N}{\sum_{j=1}^{N}{\mathbf{M}'_{ij}(p_i-\tilde p_i
)(p_j-\tilde p_j)}} \label{d2}
\end{equation}
where elements of matrix $\mathbf{M}'_{ij}$ in Eq.~(\ref{d2}) together
with the values of $\tilde p_i$ are considered as constants which
provide the most efficient selection in the form $d'^2<1$. They are
tuned for a given acceptance probability such that the proportion of
the nuclei group considered, after applying the criterion, is
maximized. Hereafter the detection rate of the nuclei group after
selection procedure is referred as the residual detection rate and in
the case of the group assigned for selection used as a measure of the
selection efficiency. In the case of selection with the help of three
image parameters the optimization consists in the determination of 9
independent constants in Eq.~(\ref{d2}). To solve this optimization
problem a general purpose routine "PIKAIA" \cite{PIKAIA} has been used.

In order to take into account the dependence of image parameters
on the impact parameter $r$ and energy $E$ of the shower and,
thus, to enhance the separation efficiency,  the image  parameters
are mapped into the space of scaled parameters
$p_{i\mathrm{sc}}=p_i/\bar p_i^{( \beta ) }( E, r )$ where $\bar
p_i^{( \beta )} ( E, r )$ is the mean value of $p_i$ for air
showers initiated by nuclei of group $\beta$ with reconstructed
values of energy and impact parameter equal to $E$ and $r$,
respectively.

In Fig.~12 and Table~\ref{ISel} we present some results
on the selection of protons and the HVH nucleus group. The
following conclusion can be drawn.

\begin{table}
\caption{Quantities related to the separation of  protons and the
HVH group by means of simultaneous application of $Width$, $Conc$
and $H_{\mathrm{max}}$ parameters.}
\begin{tabular}{|l| c|c|c|c| c|c|c|c|}\hline
Group assigned for selection &\multicolumn{4}{c|}{P}&
\multicolumn{4}{c|}{HVH}\\ \hline
 Current nucleus group &P&He&LM&HVH&P&He&LM&HVH  \\ \hline
Proportion before selection&0.39&0.27&0.12&0.22&0.39 &0.27 &0.12
&0.22\\ \hline Proportion after selection&0.89&0.11&0.00&0.00&0.00
&0.00&0.04 &0.96\\ \hline Acceptance probability&0.15&0.03&0.00
&0.00&0.00 &0.00&0.01 &0.15\\ \hline
\end{tabular}
\label{ISel}
\end{table}

\begin{itemize}
\item
The proportion of selected nuclei in the residual detection rate
grows with the reduction of the probability of acceptance of these
nuclei. This is not only true for a single image parameter but
also for any combination of such parameters.
\item
$Width$ is the most effective single image parameter. For a
fixed value of the acceptance probability of selected nuclei it
provides the   maximum contribution of these nuclei to the
residual detection rate if compared to other image parameters.
\item
The more image parameters are involved in the selection criteria
the more effective is the separation\footnote{However, our
analysis has shown that inclusion in the separation procedure of
any image parameters additional to the set of $Width$, $Conc$ and
$H_{\mathrm{max}}$ (these include $Length$ and ratio $SIZE/N_{0}$
where $N_{0}$ is the number of pixels included into analysis) does
not enhance considerably the separation efficiency.}.
\item
An overwhelming majority of selected  nuclei in the residual
detection rate (e.g. $\ge$ 90\%) can be reached only for values
$\ll 1$ of the acceptance probability of these nuclei.  This is
true even in the case of simultaneous use of three image
parameters.
\item
The simultaneous application of three image parameters and usage
of values of the acceptance probability $\kappa\le 0.20$ gives a
possibility to enrich the detection proportion to $\simeq$90\% of
selected nuclei. Under this condition a strong suppression of
other nucleus groups is possible. For example (see
Table~\ref{ISel}), using $\kappa=0.15$ and selecting the HVH
nuclei one can have 96\% of such nuclei in the residual detection
rate. The contribution of protons and $\alpha$-particles is
completely suppressed in this case; the contribution of LM nuclei
is only about 4\%.
\end{itemize}

In Table~\ref{Modes} the proportion of protons and the
HVH group in the residual
detection rate is presented for  the case of the simultaneous
application of three image parameters and for
$\kappa_\mathrm{P}=\kappa_{\mathrm{HVH}}=0.15$.
The results presented in this table correspond to different
assumptions concerning the pixel size of a multichannel camera and
the aberrations of a telescope mirror. It is seen that neither the
mirror aberrations nor the variation of the pixel size within
$0.3-0.5^{\circ}$ influence seriously the separation efficiency.

\begin{table}
\caption{The proportion of protons and HVH nuclei (assigned for
selection) in the residual  detection rate of the cell for a fixed
value of the acceptance probabilities $\kappa_{\mathrm{P}}\simeq
\kappa_{\mathrm{HVH}}\simeq 0.15$. The separation is performed
with the help of simultaneous application of $Width$, $Conc$ and
$H_{\mathrm{max}}$. Different assumptions concerning the camera
pixel size and the mirror aberrations are considered. 1 --
$PS=0.3^{\circ}$, no aberrations. 2 -- $PS=0.5^{\circ}$, no
aberrations. 3 -- $PS=0.3^{\circ}$, aberrations are included.}
\begin{tabular}{|l| c|c|c| c|c|c|}\hline
Group assigned for selection
&\multicolumn{3}{c|}{P}&\multicolumn{3}{c|}{HVH}\\ \hline
Assumptions    &1&2&3&1&2&3   \\ \hline Proportion after
selection&0.89&0.90&0.93&0.96&0.95&0.95\\ \hline
\end{tabular}
\label{Modes}
\end{table}

All results on the separation efficiency presented above
correspond to the case of the minimum available value of the
energy threshold of  the cell. In the analysis of CR energy
spectra it is important to know the dependence of this efficiency
on the air shower energy. In this relation we give in
Table~\ref{EfviaE}  the proportions  of selected nuclei (protons
and HVH ones) for different values of the hardware triggering
parameter $q_0$, i.e. for different energy thresholds
$E_{\mathrm{th}}$. It is seen that a growth of the energy of
detected air showers leads to a rapid increase of the residual
detection rate of the selected nuclei. For example, in the energy
region $\ge$100~TeV $96-99$\% of air showers surviving after
selection belong to the selected HVH nucleus group.

\begin{table}
\caption{Proportions $\eta$ of protons and HVH nuclei in the
residual detection rate via the hardware triggering parameter
$q_0$ under condition of simultaneous application of $Width$,
$Conc$ and $H_{\mathrm{max}}$ parameters and
$\kappa_{\mathrm{P}}\simeq \kappa_{\mathrm{HVH}}\simeq 0.15$.
$E_{\mathrm{th}}$ is  the effective energy threshold (TeV) of the
cell.}
\begin{tabular}{|l|c|c|c|c|c|c|c|c|c|}  \hline
  &\multicolumn{2}{c|}{P}& \multicolumn{2}{c|}{HVH}&
  &\multicolumn{2}{c|}{P}& \multicolumn{2}{c|}{HVH}  \\ \hline
 $q_0$& $E_{\mathrm{th}}$& $\eta$& $E_{\mathrm{th}}$& $\eta$&
 $q_0$& $E_{\mathrm{th}}$& $\eta$& $E_{\mathrm{th}}$& $\eta$\\ \hline
    10&       16&   0.89&       38&   0.96&
    30&       61&   0.98&      141&   0.99   \\  \hline
    20&       43&   0.91&      101&   0.96&
    60&      149&   0.99&      203&   1.00    \\  \hline
\end{tabular}
\label{EfviaE}
\end{table}

\section{Conclusion}
We suggest a new approach to study the mass composition of cosmic
rays in the energy region from several dozen TeV/nucleus up to the
knee region, i.e. up to a few PeV/nucleus, using an array of
imaging atmospheric Cherenkov telescopes with an architecture
sufficiently different from traditional IACT systems (like HEGRA,
HESS or VERITAS). This array consists of imaging  telescopes with
a relatively small mirror size ($\sim$ 10 m$^2$) separated from
each other by large distances  ($\sim$500~m) and equipped  by
multichannel cameras with a  modest pixel size ($0.3-0.5^{\circ}$)
and a  rather large viewing angle ($6-7^{\circ}$). A square cell
consisting of four telescopes  is used  to investigate basic
properties of the array.

It is shown that the cell provides considerably more effective
detection of air showers initiated by heavy nuclei compared  with
the traditional IACT systems. For example, the proportion  of
nuclei  with the charge number $Z\ge 10$ in the cell detection
rate is ($\sim$ 15\%), whereas for traditional systems this
proportion is less than 5\%.

It is found that the approach considered here could provide a
precise determination of primary energy of air showers. Its energy
resolution is within 25-35\% and depends only weakly on the air
shower energy and the atomic number of primary nucleus.

To separate air showers created by  different primary nuclei with
the IACT technique one needs image parameters sensitive to the
variation of the nucleus atomic number. We show in this study that
a number of  image parameters exhibits such a sensitivity. The
$Width$,  $Conc$  and $H_{\mathrm{max}}$ parameters are the most
effective ones.

An effective technique exploiting a criterion similar to the distance
in multiparameter space has been developed to classify air showers
detected by the cell on the basis of differences in the image
parameters. This technique allows to enrich the residual detection rate
of the cell (which includes events satisfying to the selection
criteria) up to 90\% of air showers belonging to the nucleus group
assigned for selection saving about 15-20\% of these showers.

The technique considered here could provide us with a sufficiently
good statistics of observations everywhere in the energy region
30-1000~TeV. For example, an IACT array with a modest number of
cells (16) could provide detection in the energy region above
$\sim 200$~TeV about 2000 nuclei with $Z\ge 10$ for a modest
observation time  (100~h). This number is considerably greater
than the total number of such nuclei have been accumulated until
now in the balloon and satellite born experiments.

Not having the ability of precise measurement of the primary particle
mass as in satellite or balloon borne experiments, the proposed system
nevertheless comprises capability of separation of primaries owing to
which it could be an instrument complementary to ground-based particle
detectors. To have confident results the number of predefined nuclei
groups in ground-based experiments is typically limited to 2-3 for the
cases of event by event analysis (HEGRA~\cite{HEGRA_light}, KASCADE
\cite{Vardanyan}, \cite{Roth}). The higher numbers are available only
for techniques utilizing the information on the registered spectra of
secondaries (four groups reported in~\cite{Ulrich}). According to
\cite{HEGRA_light} and \cite{Vardanyan} the proposed system would
deliver the rather higher efficiency of proton selection than that for
the HEGRA and the KASCADE for the time being. For the first project the
purity $\eta_{\mathrm{(P, He)}}=0.97$ for (P, He) nuclei group at
$\kappa_\mathrm{(P, He)}=0.3$ is reported. In our case such a value of
$\kappa$ yields $\eta \simeq 0.85$ for protons only, thus successfully
rejecting group of He nucleus which comprises $\simeq 40\%$ of the (P,
He) group. For the KASCADE $\kappa_\mathrm{(P, He)}=0.7$ corresponds to
$\eta_\mathrm{(P, He)} \simeq 0.9$. Only for protons our approach
yields $\eta_\mathrm{P} \simeq 0.65$. Comparison of our data on
selection of HVH group with that for the KASCADE (\cite{Vardanyan} and
\cite{Roth}) shows that the proposed system is inferior to the KASCADE
unless $\kappa_\mathrm{HVH} \lesssim 0.5$. At the same time for HVH and
LM groups the selection capabilities of the KASCADE deteriorates with
the decrease of primary energy \cite{Roth} so that our approach would
benefit in the sub-knee energy region.

\begin{ack}
We are grateful to A. Akhperjanian for supplying us with numerical
results concerning the mirror
aberrations of imaging telescopes.
\end{ack}


%
\newpage\clearpage

\begin{figure}[p]\label{CRSpec}
\centering
\includegraphics[width=9.7cm, angle=-90]{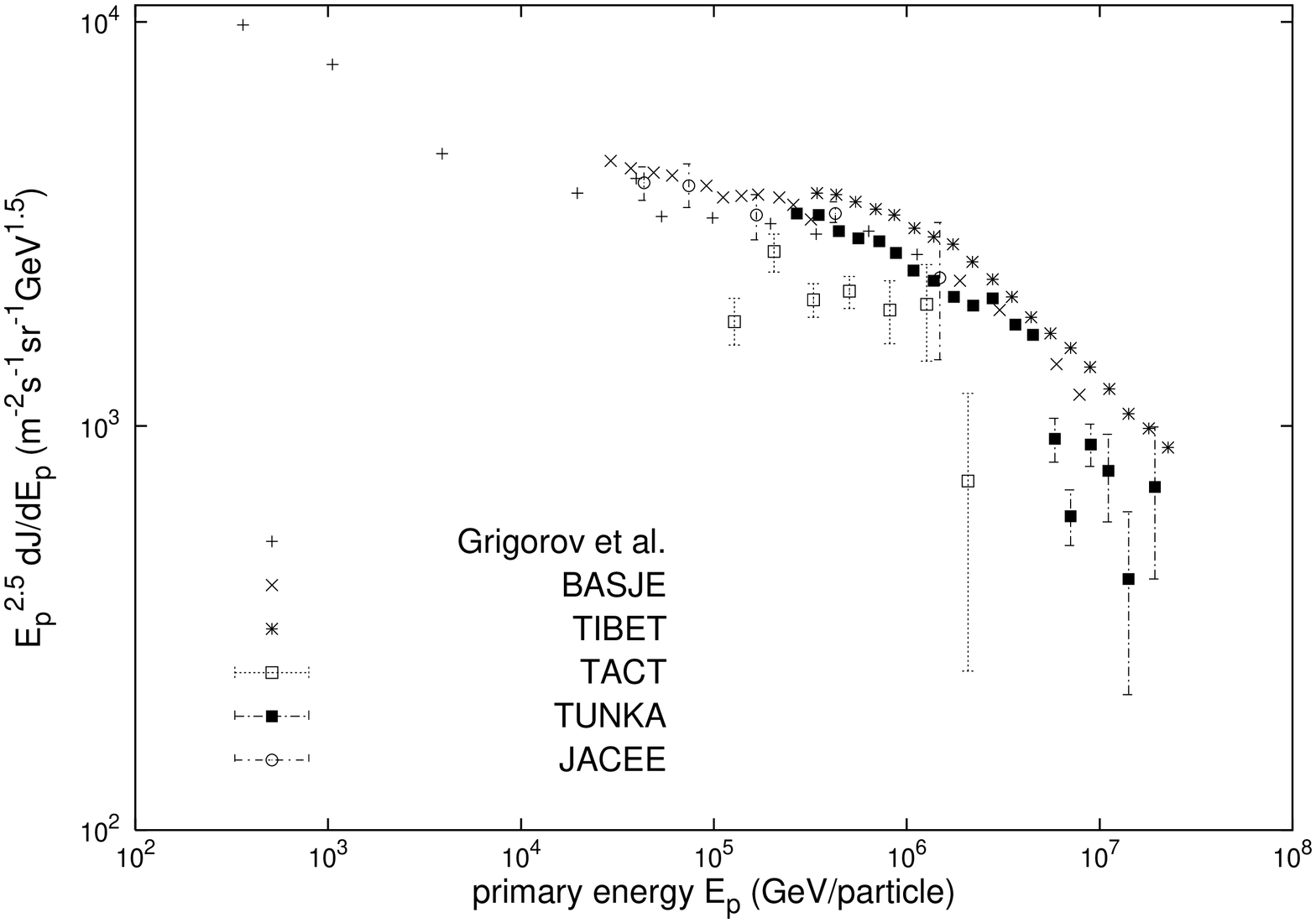}
\vspace{1cm}
\caption{The compilation of the experimental 
data on the all-particle CR flux. The figure is taken from 
Ref. \cite{Shibata} } 
\end{figure}

\newpage\clearpage
\begin{figure}\label{SviaR}
\centering
\includegraphics[width=9.7cm, angle=-90]{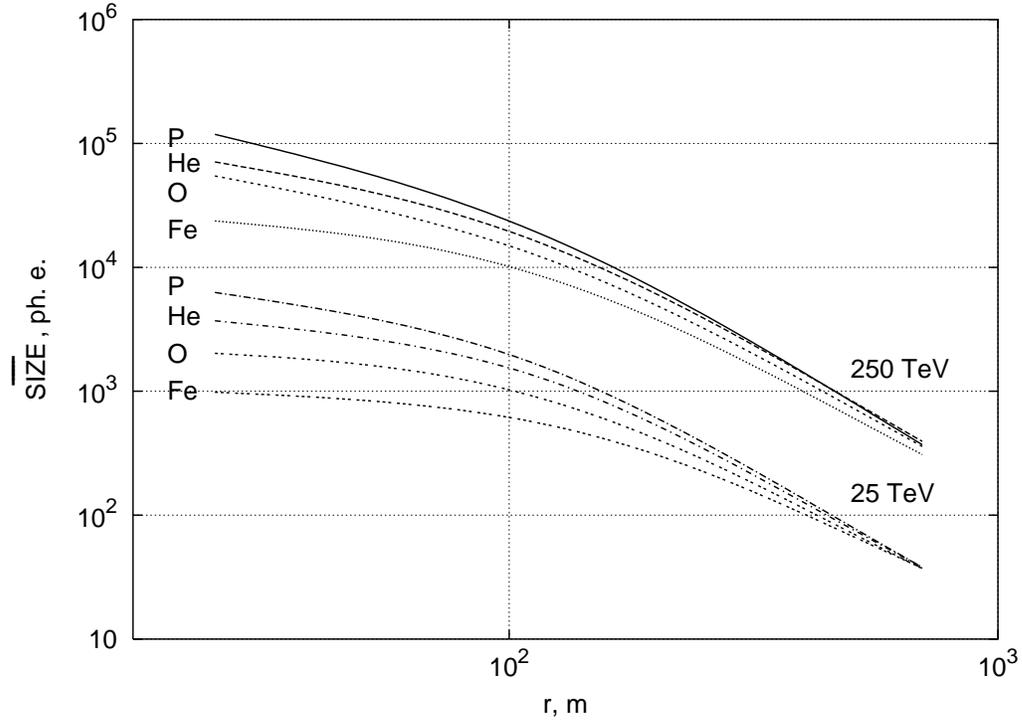}
\vspace{1cm}
\caption{The mean value of the 
total number of photoelectrons in
the image as a function of the impact parameter of air shower for
different primary energies and different primary particles
(indicated in the figure). $FoV=\infty$. No cuts on pixel
magnitudes have been applied. Contribution from the night sky
light has been neglected. }
\end{figure}

\newpage\clearpage
\begin{figure}\label{DifDetR}
\centering
\includegraphics[width=9.7cm, angle=-90]{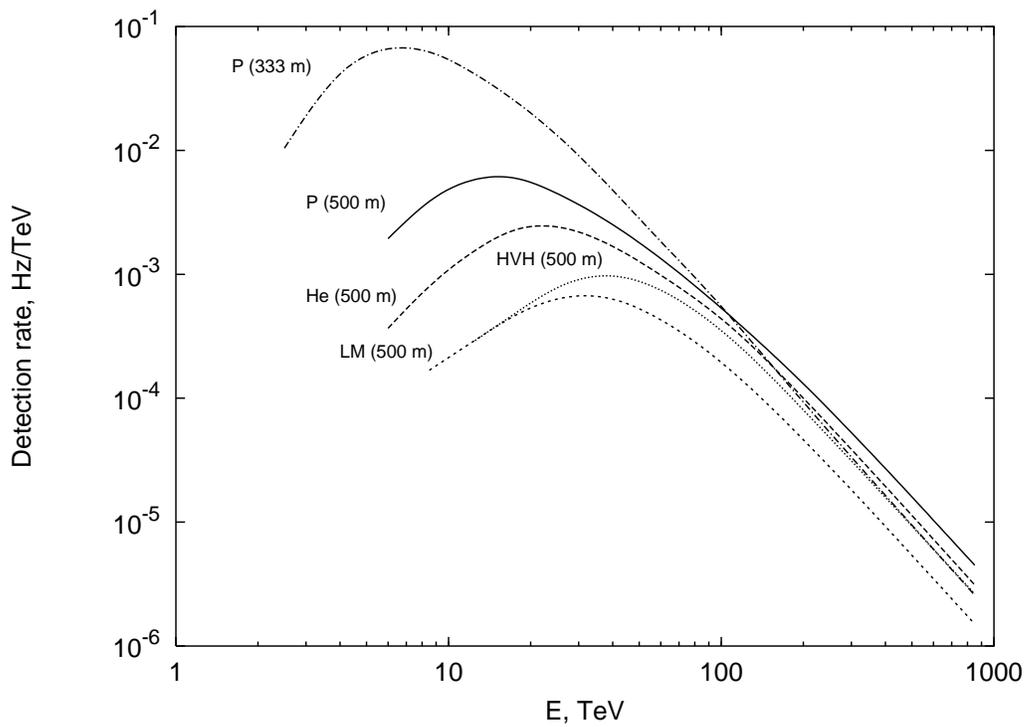}
\vspace{1cm} 
\caption{The differential detection rate of the IACT cell for
different sidelengths $l$ and different primary particles
(indicated in the figure). $FoV=6.3^o, PS=0.3^o$. $q_o=10$~ph.e.}
\end{figure}

\newpage\clearpage
\begin{figure}\label{DetRviaFoV}
\centering
\includegraphics[width=9.7cm,angle=-90]{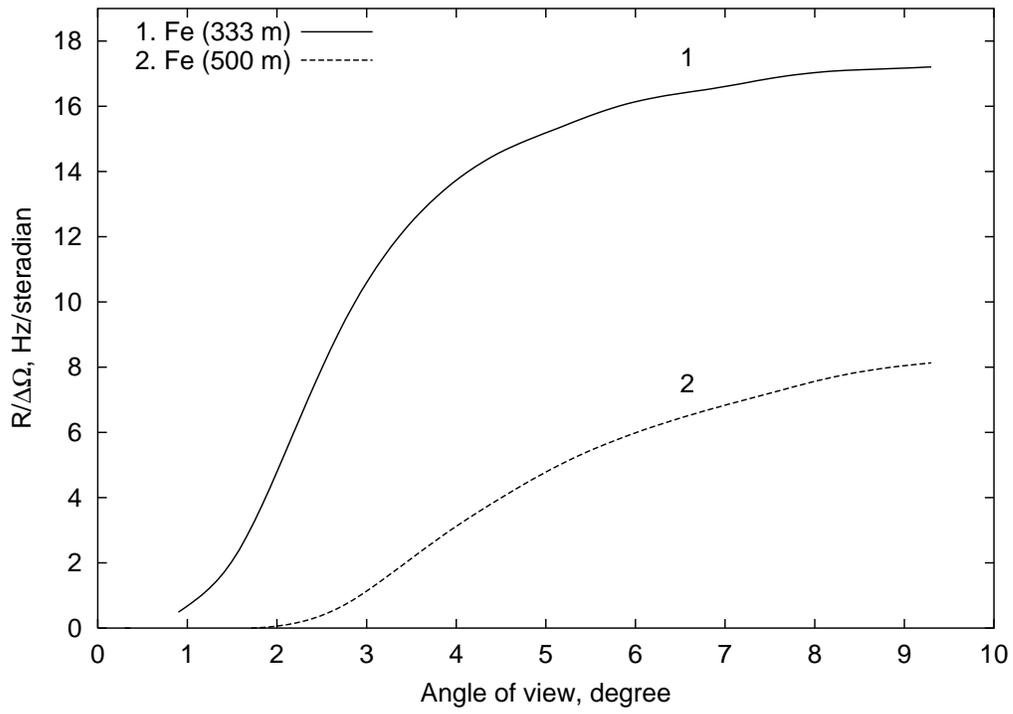}
\vspace{1cm}
\caption{The total detection rate of the cell (per one steradian
of the camera field  of view) as a function of the camera viewing
angle for different  sidelengths $l$ (indicated in the figure).
The primary iron nucleus. $q_0=$10~ph.e., $PS=0.3^o$.}
\end{figure}

\newpage\clearpage
\begin{figure}\label{RPDF}
\centering
\includegraphics[width=9.7cm,angle=-90]{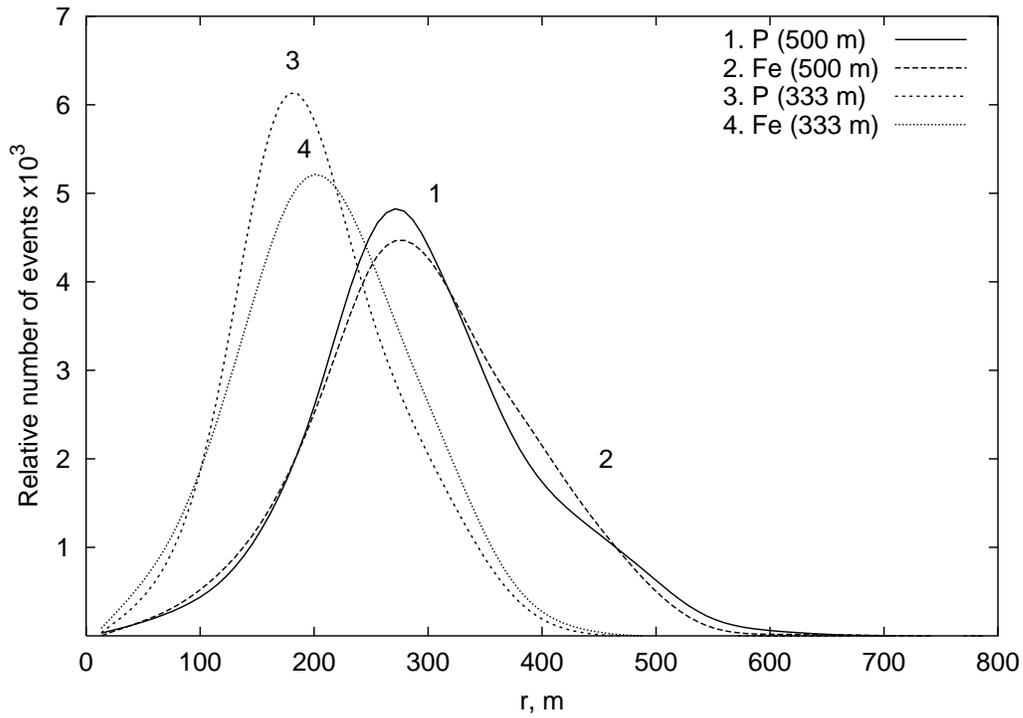}
\vspace{1cm}
\caption{The probability distributions of the air shower impact
parameter for different primary particles and different values of
the sidelength $l$.}
\end{figure}

\newpage\clearpage
\begin{figure}\label{WviaR}
\centering
\includegraphics[width=9.7cm, angle=-90]{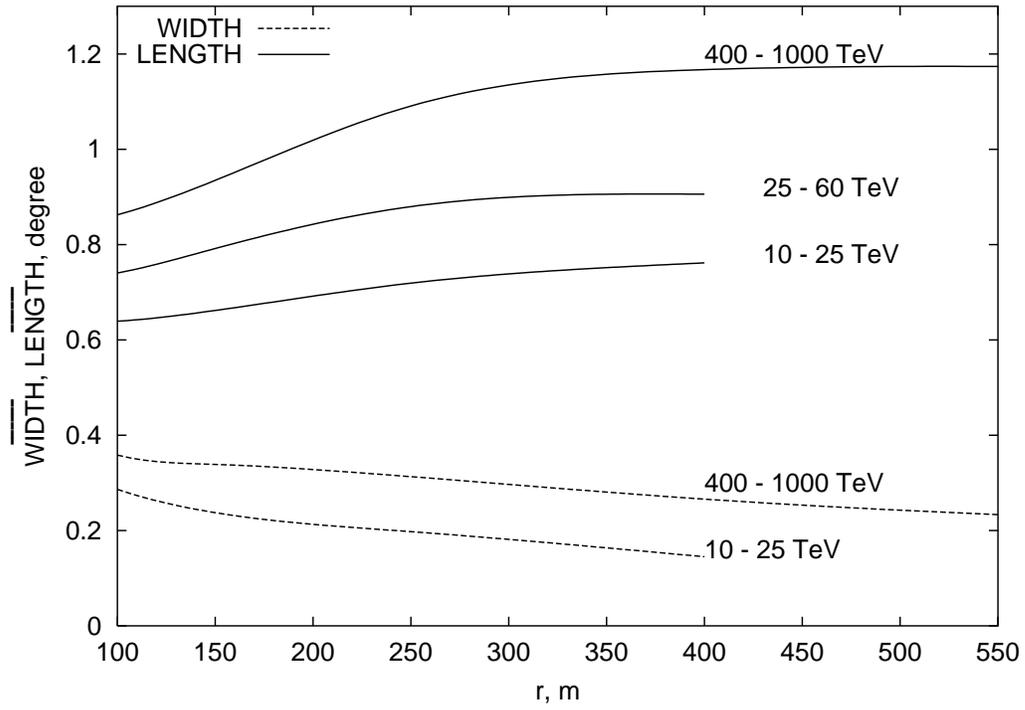}
\vspace{1cm} 
\caption{The mean values of $Width$ and $Length$ as functions of
the impact parameter for the primary proton and different
intervals of the primary energy.}
\end{figure}

\newpage\clearpage
\begin{figure}\label{WidthPDF}
\centering
\includegraphics[width=9.7cm, angle=-90]{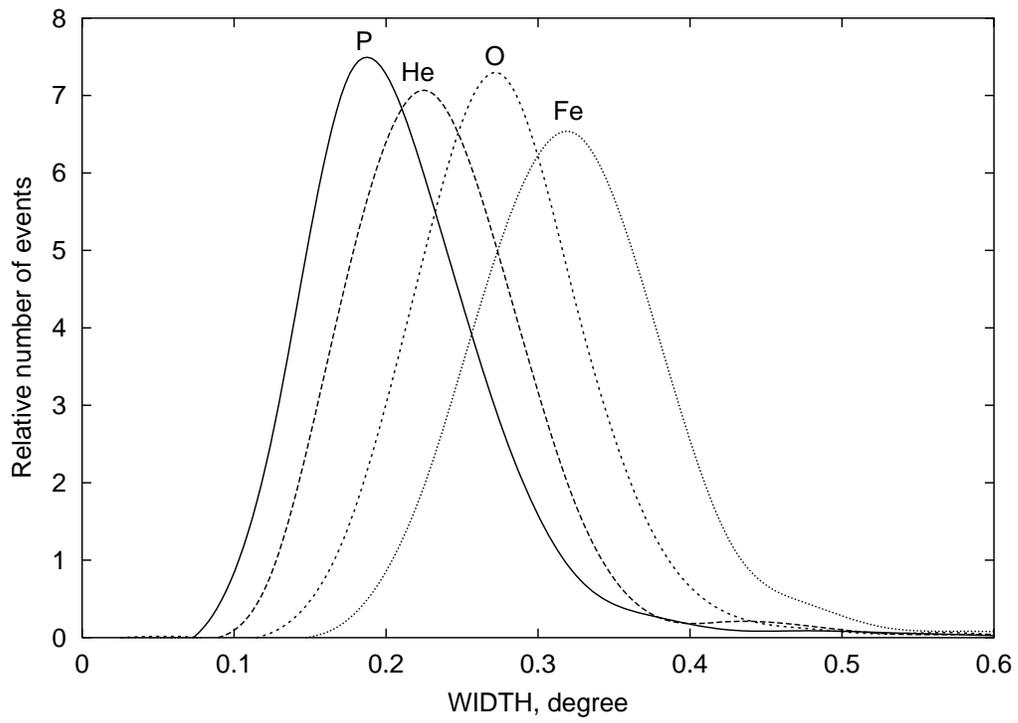}
\vspace{1cm} 
\caption{The probability distributions of $Width$ for different
primary nuclei.}
\end{figure}

\newpage\clearpage
\begin{figure}\label{HMPDF}
\centering
\includegraphics[width=9.7cm, angle=-90]{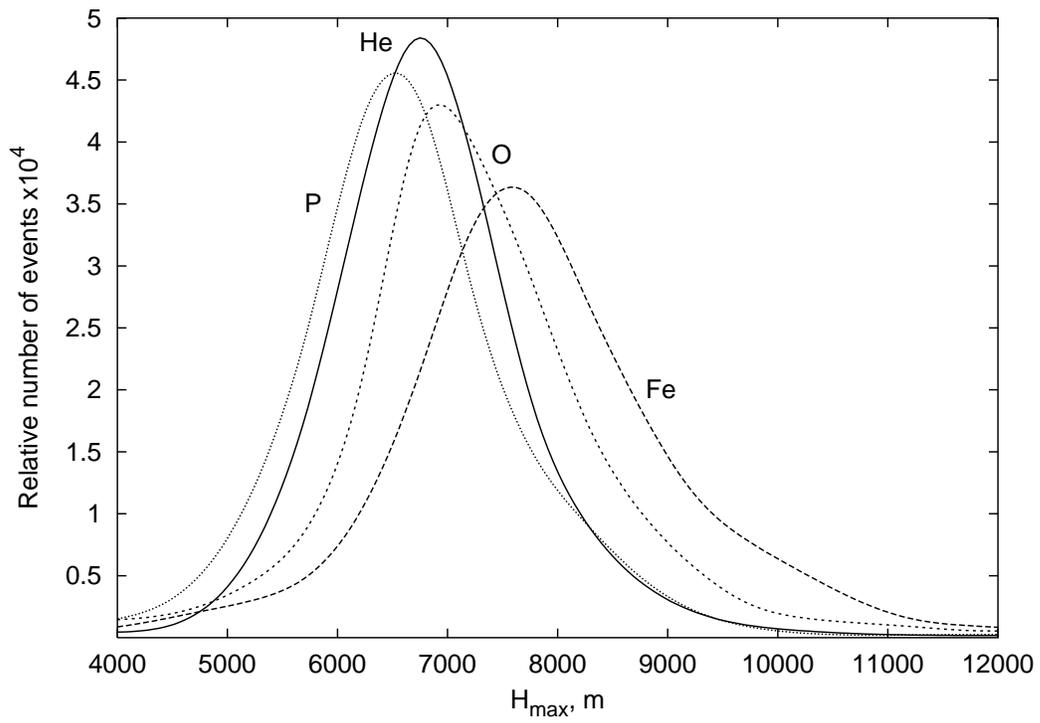}
\vspace{1cm} 
\caption{The probability distributions of $H_{max}$ for different
primary nuclei.}
\end{figure}

\newpage\clearpage
\begin{figure}\label{RErr}
\centering
\includegraphics[width=9.7cm, angle=-90]{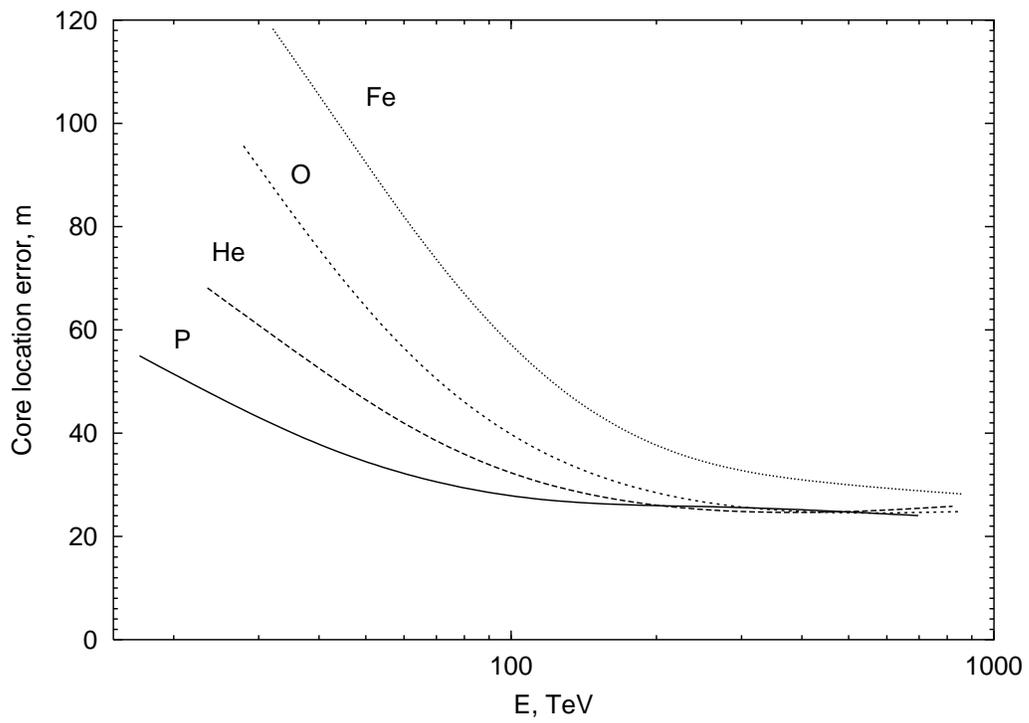}
\vspace{1cm} 
\caption{The energy dependence of the core location error for
different primary nuclei.}
\end{figure}

\newpage\clearpage
\begin{figure}\label{EErr0}
\centering
\includegraphics[width=9.7cm, angle=-90]{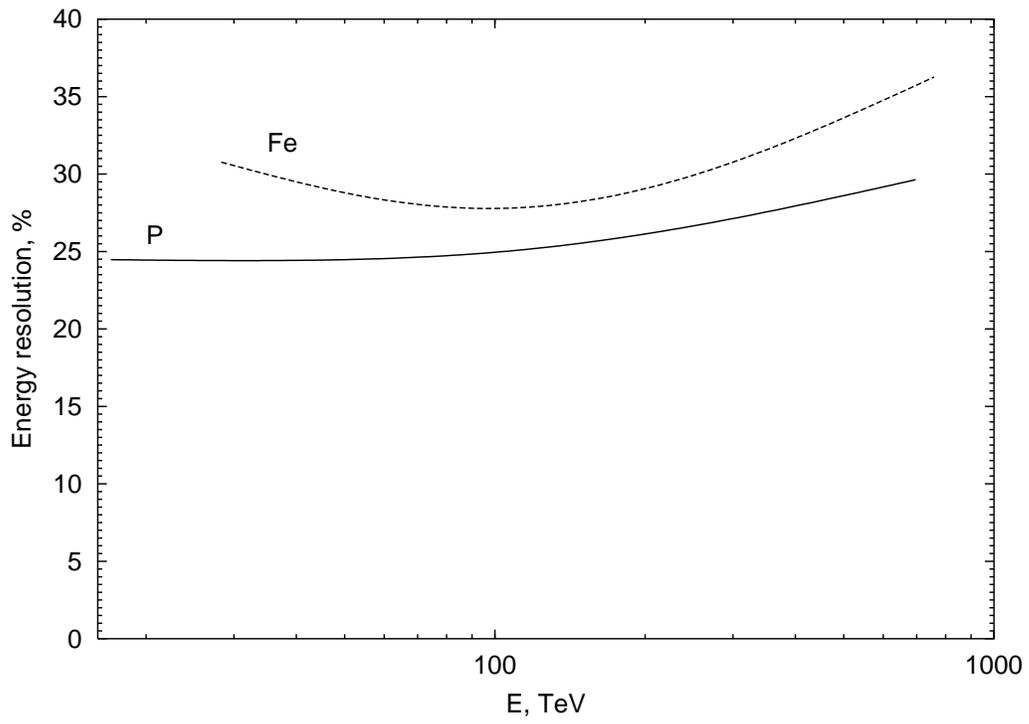}
\vspace{1cm}
\caption{The energy dependence of the cell energy resolution for
different primary nuclei.}
\end{figure}

\newpage\clearpage
\begin{figure}\label{EErr}
\centering
\includegraphics[width=9.7cm, angle=-90]{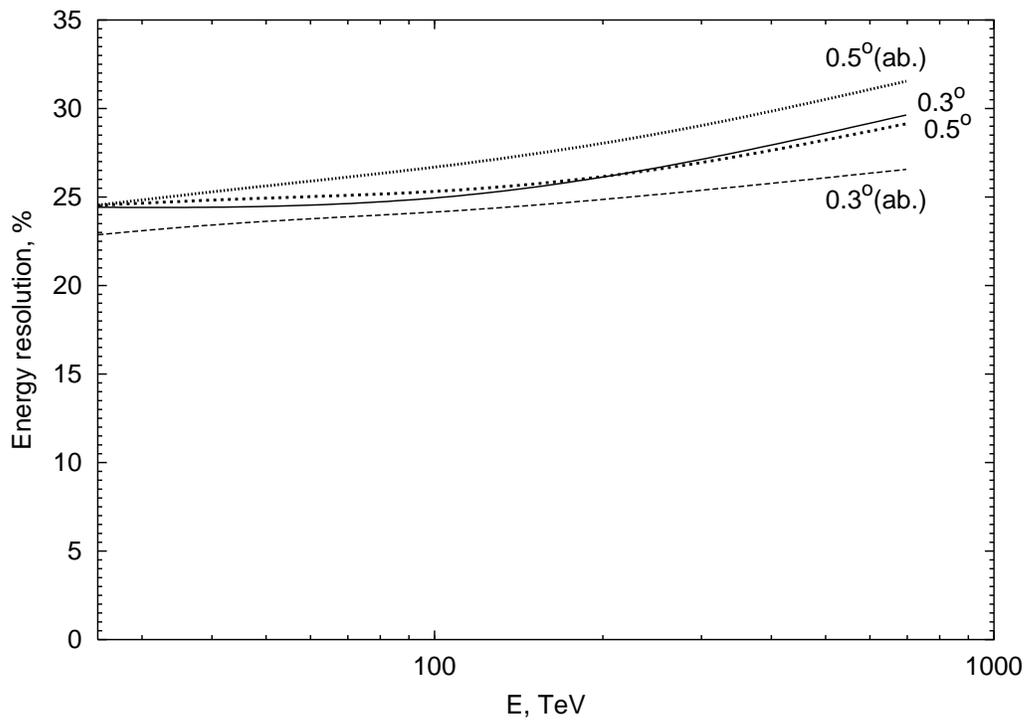}
\vspace{1cm} 
\caption{The energy dependence of the cell energy resolution for
the primary proton. Two different values of the pixel size are
considered  with taking (or not taking) into account the mirror
aberrations.}
\end{figure}

\newpage\clearpage
\begin{figure}\label{SelEf}
\centering
\includegraphics[width=8cm, angle=-90]{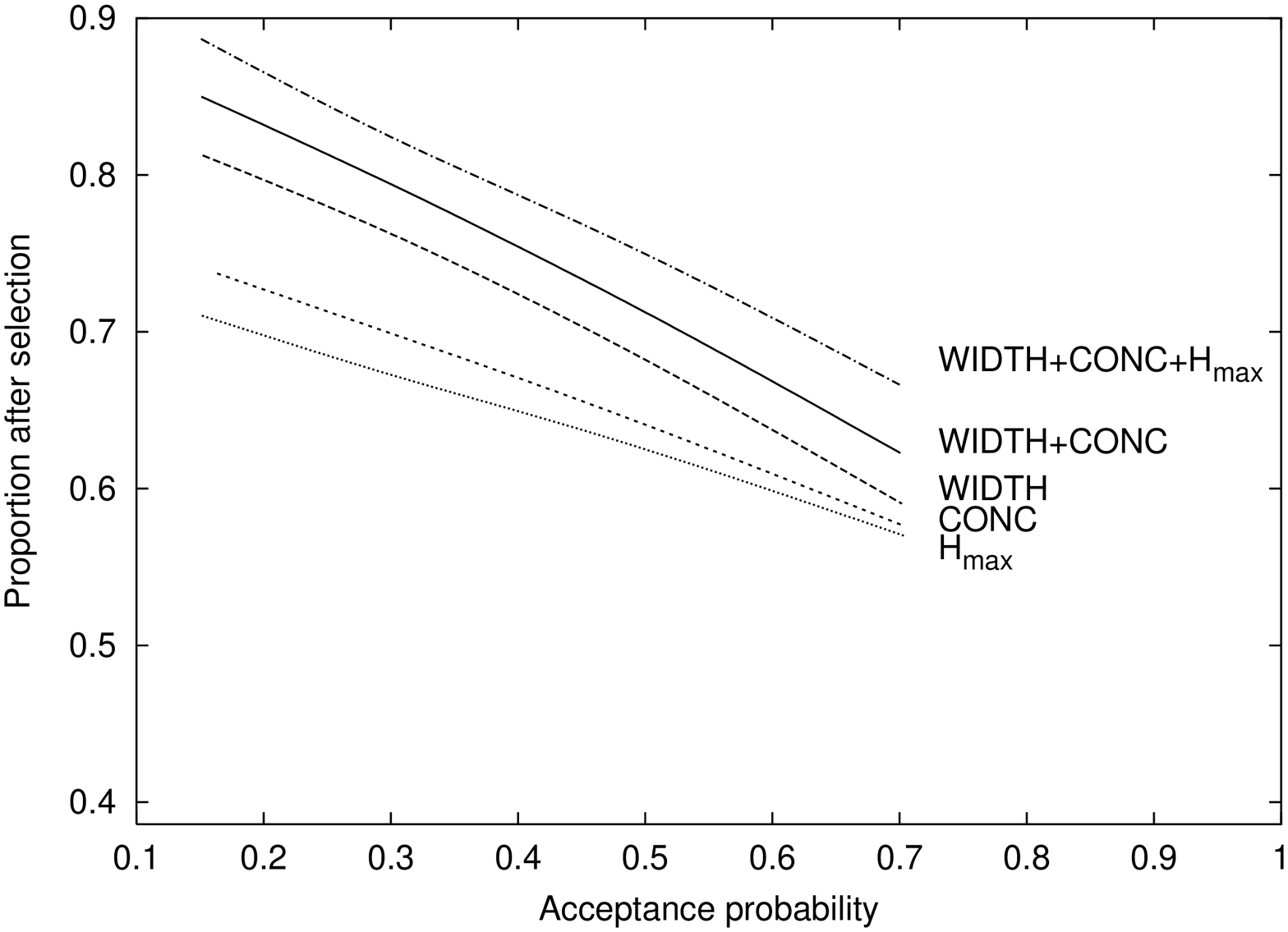}
\includegraphics[width=8cm, angle=-90]{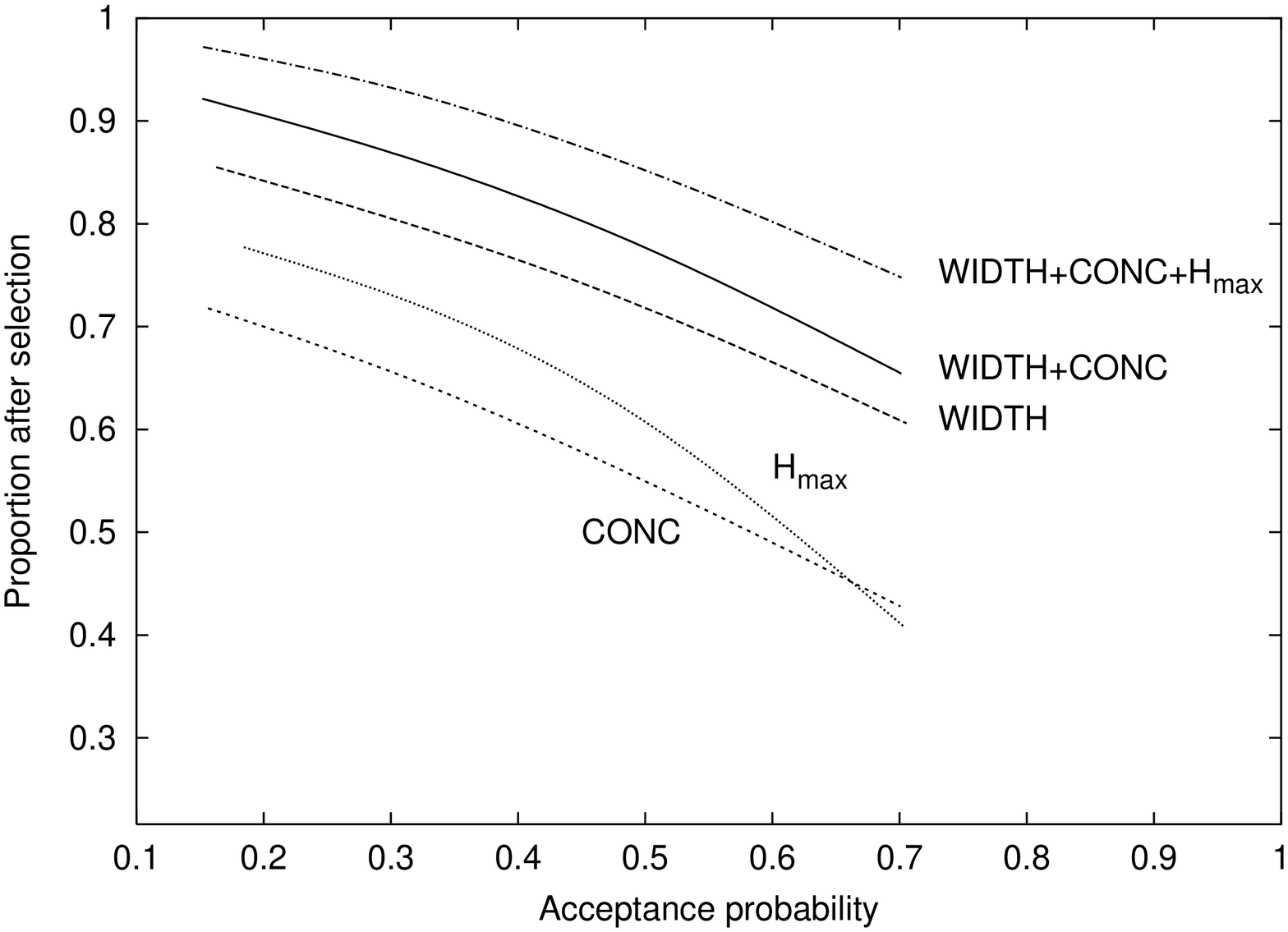}
\vspace{1cm} 
\caption{The proportion of air showers induced by CR protons
(upper panel) and HVH nuclei (bottom panel) in the residual
detection rate of the cell versus the acceptance probability of
these nuclei. Different combinations of image parameters
(indicated in the figure) are used in the separation procedure.}
\end{figure}



\begin{thebibliography}{100}
\bibitem{Cawley}
M.F. Cawley and T.C. Weekes, Exp. Astron. 6 (1996) 7.
\bibitem{Hillas}
A.M. Hillas, Space Sci. Rev. 75 (1996) 17.
\bibitem{Aharonian1}
F.A. Aharonian and C.W. Akerlof, Ann. Rev. Nucl. Part. Sci. 47
(1997) 273.
\bibitem{HEGRA}
A.K. Konopelko et al., Astropart. Phys. 10 (1999) 275.
\bibitem{HESS}
W. Hofmann In: Towards a Major Atmospheric Cherenkov Detector,
South Africa (1997) 433.
\bibitem{VERITAS}
Web-site: http://egret.sao.arizona.edu/vhegra/vhegra.html
\bibitem{Plyasheshnikov2}
A.V. Plyasheshnikov et al., J. Phys. G: Nucl. Part. Phys. 24
(1998) 653.
\bibitem{Chilingarian}
A.A. Chilingarian, Comput. Phys. Commun. 54 (1989) 381
\bibitem{PhysRev}
F.A. Aharonian et al., Phys. Rev. D 59 092003-1 (1999).
\bibitem{Shibata}
T. Shibata, Proc. of 24th ICRC, Rome, Invited, Rapporteurs and
Hightlight Papers (1995) 713.
\bibitem{Plyasheshnikov1}
A.V. Plyasheshnikov, F.A. Aharonian and H.J. V\"olk,  J. Phys. G:
Nucl. Part. Phys. 26 (2000) 183.
\bibitem{Konopelko1}
A.K. Konopelko, A.V. Plyasheshnikov,  Nucl. Instr. Meth. A 450
(2000) 419.
\bibitem{CORSIKA}
D. Heck et al., Proc. of 9th International Symposium on VHE Cosmic
Ray Interactions, Karlsruhe, Nucl. Phys. B (Proc. Suppl.) 52B
(1997) 139.
\bibitem{ALTAI_HEGRA}
F. Aharonian et al., (HEGRA Collaboration). Phys. Rev. D
59(9)(1999)2003.
\bibitem{ALTAI_WHIPPLE}
A.K. Konopelko, Whipple collaboration electronic preprint, SAO,
Tucson (1999b).
\bibitem{Wiebel}
B. Wiebel, Preprint (Wuppertal) WUB-94-08 (1994).
\bibitem{Hillas2}
A.M. Hillas, Proc. of 19th ICRC, La Jolla, 3 (1985) 445.
\bibitem{Hmax}
W. Hofmann et al., Astropart. Phys. 12 (2000) 207.
\bibitem{Aharonian3} F.A. Aharonian et al., Astropart. Phys.
6 (1997) 343.
\bibitem{Ashot}
A. Akhperjanian, private communication, 2001.
\bibitem{ESpec} G. Mohanty et al., Astropart. Phys. 9 (1998) 15.
\bibitem{PIKAIA} P. Charbonneau, Astrophys. Jour. (Supplements)
101 (1995) 309.
\bibitem{HEGRA_light} D. Horns and A. R\"ohring, Proc. of 27th ICRC, Hamburg, 1 (2001)
101.
\bibitem{Vardanyan} A.A Vardanyan, Proc. of 27th ICRC, Hamburg, 1 (2001)
67.
\bibitem{Roth} M. Roth et al., Proc. of 27th ICRC, Hamburg, 1 (2001)
88.
\bibitem{Ulrich} H. Ulrich et al., Proc. of 27th ICRC, Hamburg, 1 (2001)
97.
\end{thebibliography}
\end{document}